\documentclass[journal=apchd5,manuscript=article]{achemso}
\usepackage[hidelinks,colorlinks=true,linkcolor=blue,citecolor=blue,urlcolor=blue]{hyperref}
\usepackage{amsmath}
\usepackage[nameinlink,noabbrev]{cleveref}
\Crefname{figure}{Figure}{Figures}
\Crefname{equation}{eq}{eqs}
\creflabelformat{equation}{#2#1#3}

\usepackage{enumitem,xcolor}
\usepackage{soul}
\definecolor{Add}{RGB}{50, 200, 50}
\definecolor{Remove}{RGB}{255, 50, 50}

\usepackage[T1]{fontenc} 

\author{Amit Kumar}
\author{Sarvesh Thakur}
\author{S. K. Biswas}
\email{skbiswas@iisermohali.ac.in}

\affiliation{Bio-NanoPhotonics Laboratory, Department of Physical Sciences, Indian Institute of Science Education and Research Mohali, Knowledge City, Sector 81, SAS Nagar, Manauli PO 140306, India}

\title{An iterative method based FLC-SLM system design for forming multiple complex structures simultaneously in 3D volume with tissue}
\abbreviations{SLM, FLC, FLC-SLM, WMA-GA, 2D, 3D, NLC-SLM, SSA, CSA, PA, TM, OPC, DOPC, SBR, DMD, GG, PBR, ROI, PBS}

\begin{document}
\begin{abstract}
Complex structure formation and fast focusing of light inside or through turbid media is a challenging task due to refractive index heterogeneity, random light scattering and speckle noise formation. Here, we have proposed a weighted-mutation assisted genetic algorithm (WMA-GA) with an R-squared metric based fitness function that advances the contrast, resolution, focuses light tightly and does fast convergence for both simple and complex structure formation through the scattering media. As a compatible system with the binary WMA-GA, we have presented a fast, cost-effective, and robust iterative wavefront shaping system design with an affordable ferroelectric liquid crystal (FLC) based binary-phase spatial light modulator (SLM). The proposed wavefront shaping system design has been used to construct multiple complex hetero-structures simultaneously in 3D volume by an optimized single phase-mask. The WMA-GA and the prototype system have been validated with 120, 220, 450, and 600 grit ground glass diffusers along with 323, 588, and 852\,$\text{\textmu m}$ thick fresh chicken tissues including fluorescence in it. We have demonstrated the robustness of the proposed method to control the photon-in and photon-out from a localized fluorescent dye embedded in the tissue. The detailed results show that the proposed class of algorithm-backed integrated system converges fast with higher contrast and advances the resolution.


\end{abstract}

\section{Introduction}
Scattering of light or electromagnetic waves in living or non-living tissue and other disordered media is one of the primary challenges in biology, adaptive optics, deep tissue imaging, biomedical engineering, and is a currently active research area in biomedical imaging and medical science communities \cite{VellekoopNatPhot2010, doi:10.1021/acsphotonics.8b01354,doi:10.1021/acsphotonics.8b00948, Berto2019, doi:10.1126/sciadv.aau1338, DongWang2021, MoskNatPhot2012, OjambatiNatPhy2022, doi:10.1021/acsphotonics.0c01154, GeneNN, BonifaceNatComm2020, ConkeyNatComm2015, AriasOptica:20}. Focusing light through scattering media such as tissue has many applications in fluorescence imaging \cite{BonifaceNatComm2020, Vellekoop:10}, live cell imaging, neuron excitation/imaging \cite{doi:10.1146/annurev.neuro.051508.135540,10.1093/jnci/90.12.889}, optical trapping \cite{Cizmar2010}, and structured illumination microscopy imaging\cite{doi:10.1021/acs.analchem.1c00660}. Researchers around the globe are actively involved in overcoming the scattering problems faced in all forms of optical and other radiation-based biomedical imaging \cite{VellekoopNatPhot2010, MoskNatPhot2012, OjambatiNatPhy2022, GeneNN, BonifaceNatComm2020, ConkeyNatComm2015, AriasOptica:20,Vellekoop:10, doi:10.1146/annurev.neuro.051508.135540, 10.1093/jnci/90.12.889}. Basic understanding of physical and biological systems instructs that the inhomogeneity of the refractive index in the medium, repeated random scattering, and speckle noise due to the local interference of light cause unavoidable distortion of the wavefront. By modulating the incident wavefront, the effect of scattering can be countered to focus light inside media or through it \cite{VellekoopNatPhot2010, MoskNatPhot2012, OjambatiNatPhy2022, GeneNN}. Focusing light inside or through scattering media using spatial light modulator (SLM) based wavefront shaping was initially demonstrated experimentally by Vellekoop \textit{et al.} in 2007 \cite{Vellekoop:07}. In their work, an iterative algorithm called step-wise sequential algorithm (SSA) was used to optimize the phase mask, which was displayed on the nematic liquid crystal SLM (NLC-SLM). Later, the same group introduced the optimization of phase masks using iterative algorithms called continuous sequential algorithm (CSA) and partitioning algorithm (PA) in 2008\cite{Vellekoop2008}. Popoff \textit{et al.} in 2010 \cite{PhysRevLett.104.100601}, introduced the transmission matrix (TM) based approach as an alternative algorithm to focus light through the scattering media. Further, a number of research groups demonstrated focusing light through the stationary scattering media \cite{VellekoopNatPhot2010, MoskNatPhot2012, Vellekoop:10, PhysRevLett.104.100601, Popoff2010, Cui:11}. In 2012, Stockbridge \textit{et al.}\cite{Stockbridge:12} introduced focusing light through \textit{ex-vivo} chicken tissue which was considered as a dynamic scattering media. Imaging through scattering media using the optical phase conjugation (OPC) approach was shown by Yaqoob \textit{et al.} in 2008 \cite{Yaqoob2008}. Further, the digital optical phase conjugation (DOPC) approach was experimentally demonstrated by Cui \textit{et al.} \cite{Cui:10}, and Hsieh \textit{et al.} in 2010 \cite{Hsieh:10}.

During 2007 to 2012, wavefront shaping was mostly carried out using iterative approaches (SSA, CSA, PA), TM approach, and DOPC approach \cite{VellekoopNatPhot2010,MoskNatPhot2012, Vellekoop:07, Vellekoop2008, PhysRevLett.104.100601, Popoff2010,Cui:11, Stockbridge:12, Yaqoob2008, Cui:10, Hsieh:10}. In the presence of the high amount of environmental and instrumental noise, CSA and SSA have slow convergence because it is difficult to detect the variation in feedback signal due to mode-by-mode modulation, which results in initial measurement errors \cite{Conkey:12}. The rising computational power in the last decade has enabled the use of machine learning and advanced metaheuristic algorithms. It was found that evolution-inspired iterative optimization algorithms like genetic algorithm (GA) are well suited for the problem because of the huge solution space for the possible phase masks, and this was first demonstrated by Conkey \textit{et al.} in 2012 \cite{Conkey:12}. Relevant studies have shown that genetic algorithms perform better in terms of enhancement compared to the previously introduced iterative approaches (CSA, SSA, PA) and TM approach, even in highly noisy environments \cite{Conkey:12, 45, 57}.

Further improvements in the realm of the genetic algorithm were done by introducing micro-genetic algorithm \cite{45}, genetic algorithm with signal-to-background ratio (SBR) discriminant \cite{47}, genetic algorithm with interleaved segment correction \cite{50} and four-element division GA \cite{57}. Other iterative algorithms such as particle swarm optimization \cite{55, 54}, gradient-assisted focusing \cite{Zhao:21}, and neural networks \cite{59} have also been introduced recently. Furthermore, neural network was combined with GA by Luo \textit{et al.} in 2020 \cite{GeneNN}. These recent research findings made GA the \textit{state-of-the-art} algorithm for wavefront shaping.

However, all of the above algorithms were demonstrated using either a nematic liquid crystal SLM (NLC-SLM) or a digital micro-mirror device (DMD), and the realm of iterative binary phase modulation with ferroelectric liquid crystal SLM (FLC-SLM) is still unexplored with well-suited evolution based algorithms. Despite the introduction of different types of algorithms, essential advanced hardware such as a fast camera, high resolution phase-only SLM (NLC-SLM), or fast switching amplitude modulator (DMD) are still out of reach to most of the research groups due to lack of cost-effectiveness of these instruments.

Usually, digital micro-mirror devices (DMDs) have been used for fast focusing because they have a faster refresh rate ($\sim 23\, \mathrm{kHz}$) \cite{ConkeyDMD:12, Akbulut:11} and low latency. On the other hand, NLC-SLMs have high latency and low frame rate ($\sim 60\, \mathrm{Hz}$)\cite{57, 50, Zhao:21, Liu:17}. However, DMDs can only achieve binary amplitude modulation, which reduces the enhancement factor compared to the phase modulation achieved by either binary FLC-SLMs (having 2 discrete phase levels) or NLC-SLMs (having 256 discrete phase levels)\cite{Liu:17}. The theoretical enhancement factor of binary phase modulation with FLC-SLM is double as compared to DMD\cite{Liu:17, VellekoopReview:15}. Furthermore, DMD based experimental setups are significantly complex. Due to its oblique reflection sensitivity, the alignment is difficult, and DMDs cannot be used with high-intensity pulsed lasers \cite{Wang:15, Tay:14}. For NLC-SLM, mandatory phase calibration is required, whereas FLC-SLM does not require any kind of phase calibration. FLC-SLM is faster than NLC-SLM as it operates in binary mode. In summary, FLC-SLMs are a cost-effective alternative, provide fast binary phase modulation, do not require phase calibration, and provide more enhancement compared to DMDs. The use of FLC-SLM for focusing light in scattering media has been shown using DOPC based wavefront shaping technique\cite{Liu:17}. However, DOPC techniques have some unavoidable drawbacks. The first is that the camera pixels and the SLM pixels must be in a near-perfect match which makes alignments far more challenging \cite{Cui:10, VellekoopBinaryDOPC:12}. Furthermore, the SLM and camera have to be at the exact mirror conjugate plane \cite{Cui:10, VellekoopBinaryDOPC:12}.

Focusing light inside or through scattering media is a result of a synergy between the feedback algorithm and the hardware. In the genetic algorithms, optimization in the crossover and mutation with the fitness function has not been explored extensively. It has been observed that the weighted mutation plays an influential role in the evolution-assisted feedback loop for advancing the solution and convergence. In this context, we have developed a weighted-mutation assisted genetic algorithm (WMA-GA) for optimizing the phase mask for fast convergence and advancing contrast. A detailed derivation of the proposed WMA-GA has been presented with fundamental evolutionary theory, sampling theorem and set theory. To construct complex structures at high resolution, we have introduced an R-squared metric based fitness function in the proposed WMA-GA. Results obtained with the proposed method are compared to the standard GA\cite{Conkey:12}. In addition, Ferro-electric liquid crystals in FLC-SLM have a pixel switching response time of 40\,$\text{\textmu s}$ with a refresh rate of up to 4.5\,$\mathrm{kHz}$ at the present \cite{Park:20}. The high refresh rate, the high-speed pixel switching time and the proposed algorithm-compatible binary phase features of FLC-SLM can be used to advance the contrast in fewer iterations with the proposed method. A cost-effective wavefront shaping system has been designed with FLC-SLM and dual cameras to construct multiple complex structures at different depths simultaneously in 3D volume using an optimized single phase-mask. We have validated the algorithm and the designed system with 120, 220, 450, and 600 grit ground glass (GG) diffusers along with 323, 588, and 852\,$\text{\textmu m}$ thick fresh \textit{ex-vivo} chicken tissues including fluorescence in it. The fluorescence emission photons of light (680-710\,nm) have been enhanced inside the tissue by shaping the wavefront of exciting laser light of wavelength 633\,nm and controlled the photons penetrating in and out of the tissue. Multiple complex light structures formation will find new applications in real-time 3D holographic display\cite{YuNatPhot2017, Tran2019}, photo-thermal imaging and therapy, dosimetry, fluorescence imaging \cite{BonifaceNatComm2020, Vellekoop:10}, fast light sheet microscopy \cite{LSMYang2019, Shi2022, Dean2022}, photoacoustic microscopy imaging\cite{Hazan2022} and structure illumination microscopy\cite{doi:10.1021/acs.analchem.1c00660}.

\section{Results and Discussion}
\subsection{Principle of Proposed WMA-GA}
\begin{figure}[!ht]
\centering\includegraphics[width =0.7 \linewidth]{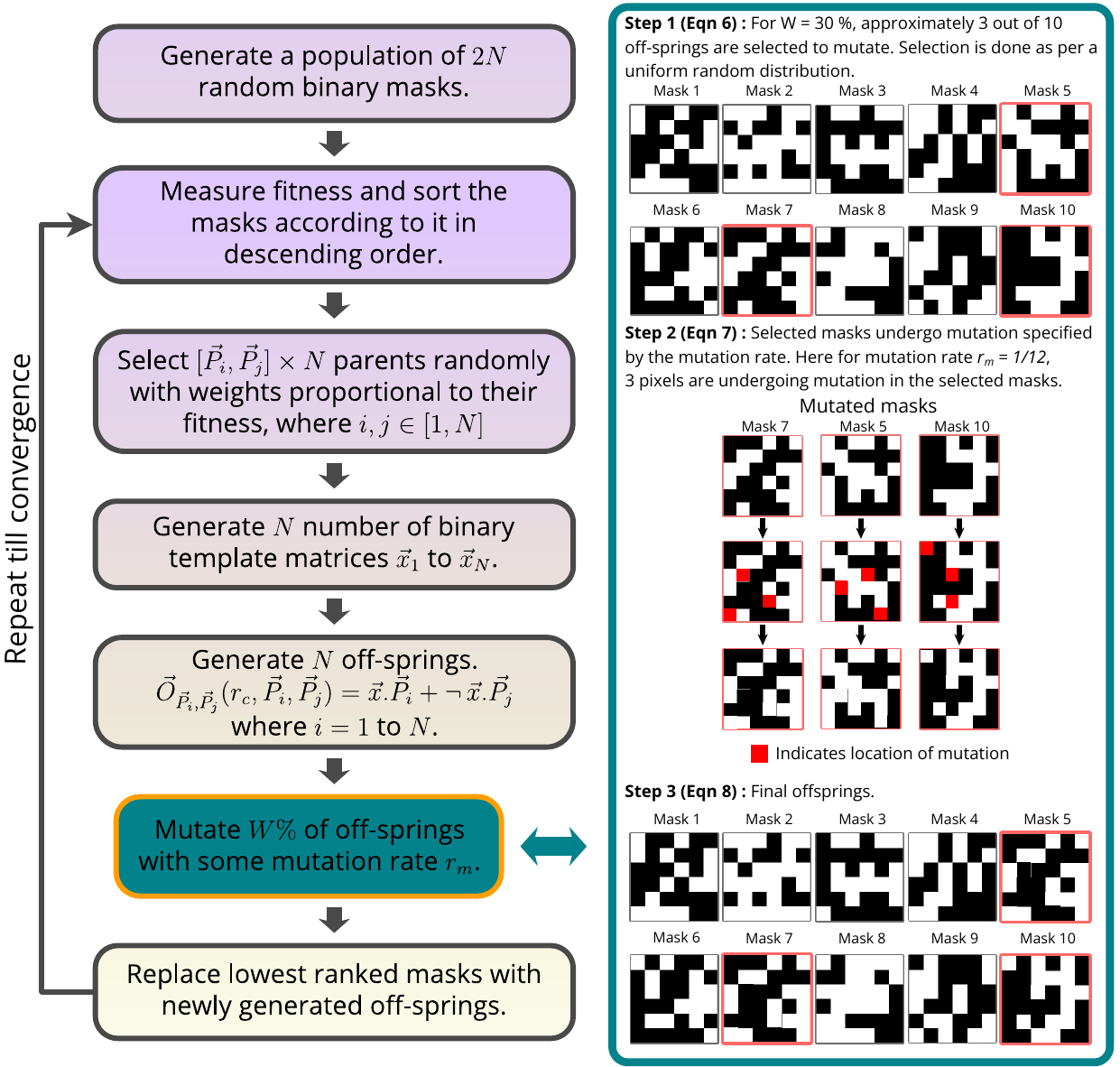}
\caption{\textbf{Flowchart of WMA-GA.} The detailed step-by-step flowchart for the working principle of WMA-GA, along with the demonstration of its weighted mutation process in the population (on the right side).}
\label{fig:flowchart_fig}
\end{figure}

The principle of the proposed WMA-GA with detailed computational steps is shown in the flowchart (\Cref{fig:flowchart_fig}). The formulation of WMA-GA, as described in the flowchart, has been derived by a set of mathematical equations (\Crefrange{eq3}{eq8}). The theoretical background and the development of the proposed WMA-GA are described in the following paragraph.\\
If an incoming light with field $E(d)$ transmits through an optical scattering media of transmission function $T(d,d^{'})$, then the output complex field can be written as $E(d^{'})$= $\sum_{d}T(d,d^{'}) E(d)$. A transmission matrix $T$ of dimensions $M \times N$ models the wavefront scattering through disordered media. Here, $T$ is generated by a Gaussian complex random matrix. The equation for the calculation of output modes $M$ can be written as \cite{Vellekoop:07,Conkey:12};

\begin{equation}E_m = \sum_{n}^{N} t_{mn} A_n e^{i{\phi}_n}
\label{eq1}
\end{equation}

Where $A_n$ and ${\phi}_n$ are the amplitude and phase of the input mode ($n$), respectively, and $t_{mn}$ is a particular element of transmission matrix $T$. The WMA-GA or standard GA starts with a population of random masks which undergo crossover followed by a mutation ($r_m$) over iterations until the convergence criteria is met \cite{Conkey:12}. The standard GA mainly follows selection and crossover operators, while the mutation operator has only secondary significance. 
Crossover being the main operator, the initial pool of solutions should be large enough to ensure a large diversity which slows the computation process. In this process, as iterations progress, diversity in the population decreases, and as a result, the chance of premature convergence increases. For representing the complete output mode vector $\vec{E}_{out}$, the input mode vector $\vec{E}_{in}$ can be written as a function of various evolution control parameters ($e$) as;
\begin{equation}
\vec{E}_{out} = T \vec{E}_{in}({e})
\label{eq3}
\end{equation}

Here, the evolution controlling parameters (e) can be mutation, crossover, an optimized weighted mutation, and population size. In \Cref{eq3}, $E_{out}$ is the complex field of the output mode, and its amplitude is chosen as $A_n=1/\sqrt{N}$. Therefore, the intensity ($I_m$) at a particular output mode at the camera with the added noise can be written as;

\begin{equation}I_m = \frac{1}{N}\Bigg{|}{ \sum_{n}^{N} t_{mn} e^{i{\phi}_n}}\Bigg{|}^2 +\, \delta \quad \text{where} \quad 
\delta = \frac{\Gamma \%}{100} \times \mathcal{N}(\mu,\,\sigma)\, I_o
\label{eq2}
\end{equation}

Here, a noise $\delta$ is added to mimic the experimental environment. $\Gamma$ represents the added noise percentage with respect to the initial average intensity $I_o$. $\mathcal{N}(\mu,\sigma)$ represents a random number generated from a normal distribution with mean $\mu$ and standard deviation $\sigma$. 

It has been observed in the case of binary phase modulation, that the pattern of crossover ($r_c$) and weighted mutation ($W_m$) optimizes the global as well as local solution. To incorporate diversity and, to escape from the premature local convergence, a biased probability based crossover and a weighted mutation ($W_m$) based local subspace perturbation is carried out to advance the global as well as local solution refinement without jumping to some other far subspace.
To explore the weighting factor in the WMA-GA, we have represented a statistical sampling function for a pattern of either crossover or mutation or both together. The resultant output mode vector with crossover and weighted mutation ($W_m$) can be written as;
\begin{equation}
\vec{E}_{out} = T \vec{E}_{in}(r_c, r_m, W_m)
\label{eq4}
\end{equation}The output mode of the \cref{eq4} depends on the weighted mutation, and the proposed algorithm is termed as `weighted-mutation assisted genetic algorithm (WMA-GA)'. To generate the mask for the next iteration, a particular off-spring $\vec{{O}}_{\vec{P}_i,\vec{P}_j}(r_c,\vec{P}_i,\vec{P}_j)$ with parents $\vec{P_i}$ and $\vec{P_j}$ is generated by crossover with a random binary vector $\vec{x}$ and its conjugate $\neg\,\vec{x}$ respectively. This particular off-spring or phase mask can be written as;
\begin{equation}
\vec{{O}}_{\vec{P}_i,\vec{P}_j}(r_c,\vec{P}_i,\vec{P}_j) = \vec{x}.\vec{P_i}+ \neg \, \vec{x}.\vec{P_j} \label{eq5}
\end{equation}

Where two parents $\vec{P}_i$ and $\vec{P}_j$ are selected with a biased probability towards a high fitness score. The descending order of the phase masks ranked according to their fitness score is used for parents selection. To generate a full set of off-springs $\mathcal{O}$, \Cref{eq5} has been followed for all the selected parents, where $\vec{{O}}_{\vec{P}_i,\vec{P}_j} \subseteq {\mathcal{O}}$. In the WMA-GA, every off-spring has not gone through mutation, but certain off-springs are selected by Bernoulli sampling\cite{ross98} $\mathcal{B}$ from the entire set $\mathcal{O}$ with probability $W_m$ for introducing the mutation. The resultant subset of selected off-springs is denoted by $O^w$, and can be written as;
\begin{equation}
{{O}}^{w} = {\mathcal{B}(\mathcal{O},W_m) \subseteq \mathcal{O}}
\label{eq6}
\end{equation}

Where, $O^{w} \subseteq O$, $W_m \in [0,1]$. Now the few sampled off-spring $\vec{O}^{w}_i$ are gone through mutation, and it can be expressed as;
\begin{equation}
\vec{O}_{M_i}^{w}(r_m,\vec{z}_i \,) = (\vec{O}^{w}_i \cdot \neg \, {\vec{z}_i \, (r_m)})+ ({\vec{z}_i \, (r_m)} \cdot \neg \, \vec{O}^{w}_i) \label{eq7}
\end{equation}

Where, $\vec{z}_i \, (r_m)$ is a biased random binary vector [0,1] which is used to do the mutation of selected off-spring $\vec{O}^{w}_i$ with the current mutation rate $r_m$. Now, the new subset of off-springs or phase masks that have gone through mutation is reunited with the rest of the non-mutated phase masks, and it is defined as;
\begin{equation}
 {O}_{new} = [\mathcal{O} \, \cap \, \neg \, O^w] \, \cup \, O_{M}^{w}(r_m,z)
\label{eq8}
\end{equation}

The newly formed set of masks ${{O}}_{new}$ is again passed on to the SLM to measure the fitness. The above mathematical operations have been implemented in the simulation together with the experiment, and the results are discussed in the following sections.\\

\begin{figure}[h!]
\centering\includegraphics[width =0.95 \linewidth]{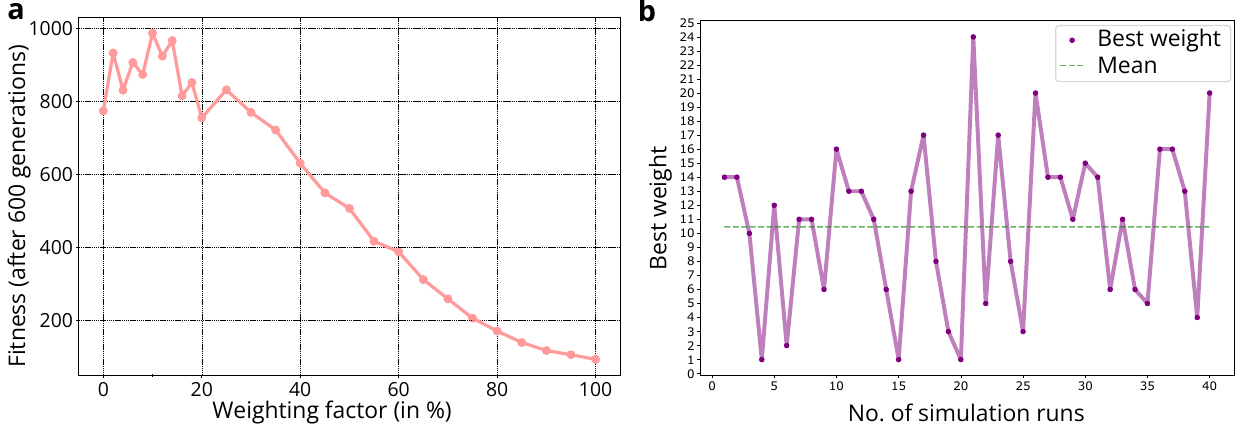}
\caption{\textbf{a.} shows the weighting factor vs. enhancement plot for WMA-GA. The comparison of fitness enhancement after 600 iterations is shown for different values of weighting factor $W\%$. \textbf{b.} shows the variation of the weighting factor with multiple runs. The best value of W varies between 0 and 25, and the mean W is found to be $\sim$10\% for the proposed WMA-GA.}
\label{fig:fig2}
\end{figure}

\begin{figure}[h!]
\centering\includegraphics[width = 0.6 \linewidth]{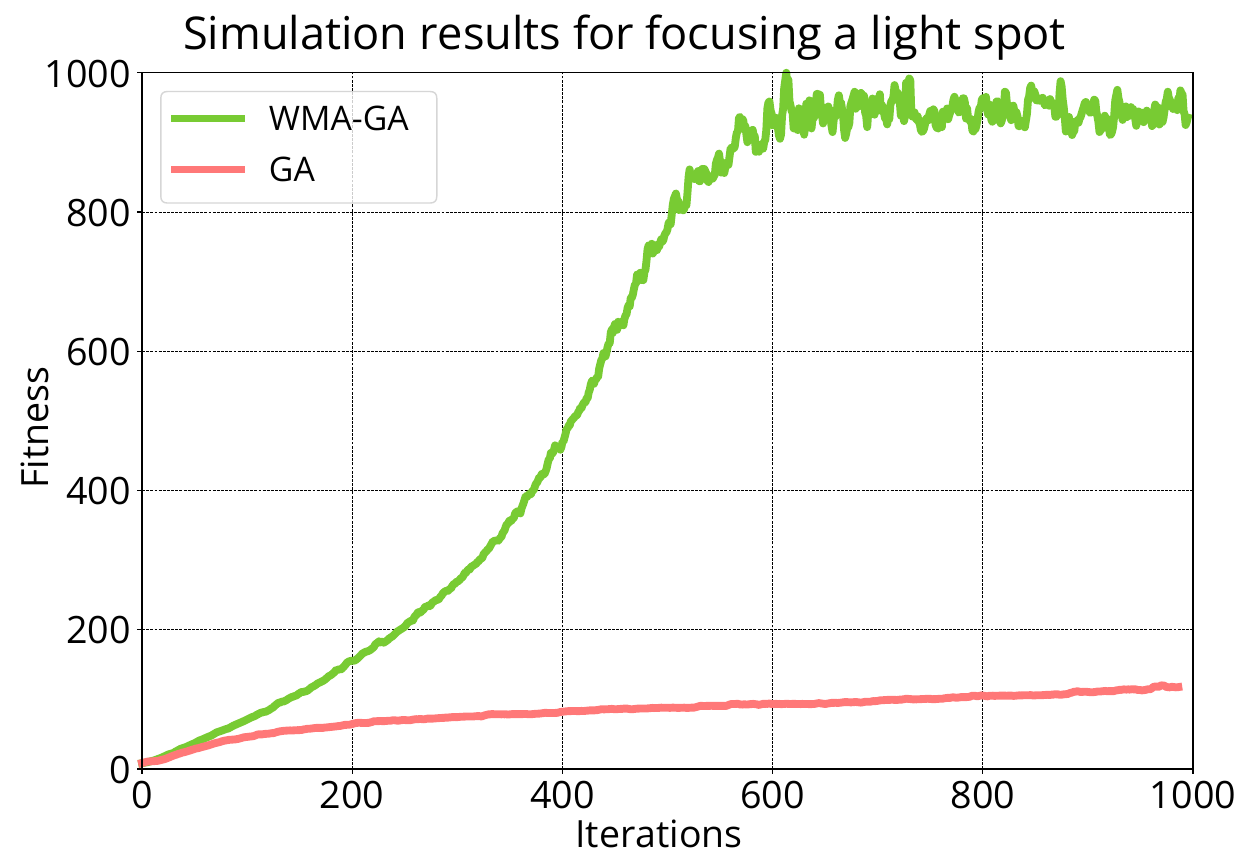}
\caption{\textbf{Simulation results for focusing a light spot.} The figure shows the progress of fitness score v/s number of iterations for focusing light using WMA-GA and standard GA.}
\label{fig:fig3}
\end{figure}

\begin{figure}[h!]
\centering\includegraphics[width = 0.8 \linewidth]{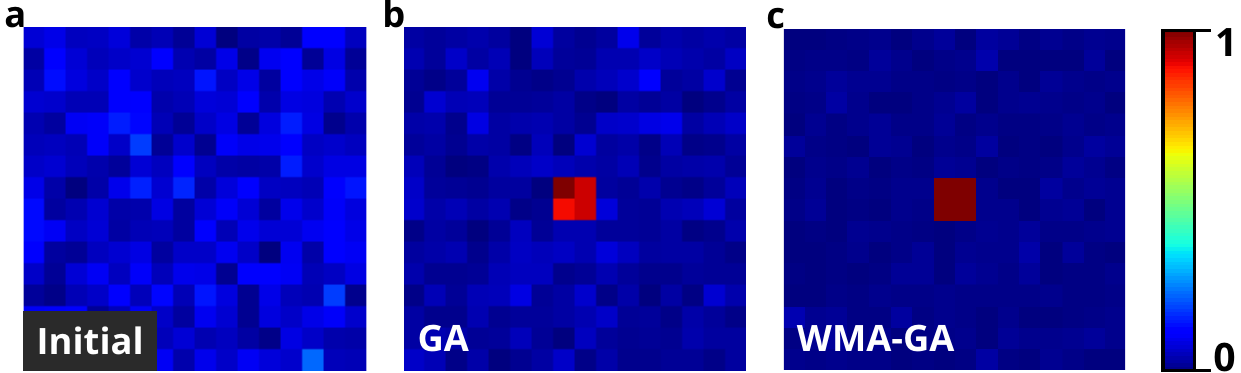}
\caption{\textbf{Simulation results for focusing light through scattering media.} Where \textbf{a.} shows the initial intensity distribution, \textbf{b.} shows the focused light spot using standard GA, and \textbf{c.} shows the focused light spot with proposed WMA-GA. It is visible that standard GA is not able to suppress background intensity compared to WMA-GA. The noise percentage added to the mask has been 30\% of the initial average intensity.}
\label{fig:fig4}
\end{figure}

\subsection{Simulation Model Analysis for Verifying the Performance of WMA-GA} 
The proposed algorithm has been tested in the simulation model. \Cref{fig:fig2}a shows the comparison of fitness score enhancement with different weighting factors varying from 0\% to 100\% for focusing the light tightly at a spot. Results are compared between the proposed WMA-GA and the standard GA in the simulations and experiments. It shows that the WMA-GA performed exceptionally good in terms of contrast enhancement and convergence speed compared to the standard GA.

The proposed algorithm has been studied extensively for the impact of various experimental conditions, such as different noise levels, mutation rates, scattering effects and input modes. Detailed studies of the proposed algorithm under various conditions have been provided in Figures S2--S8 of the Supporting Information. Based on these statistical analyses, we have obtained an optimized weighting factor (W) supported by the statistical estimate from 40 cycles. (Please see \Cref{fig:fig2}b).

The range of `W's with the best fitness value of the cost function has been observed between 0\% and 25\% with an average value of 10.4\%, where the standard deviation of the mean value of W was found to be 5.8. \Cref{fig:fig3} shows the progress of fitness scores against iterations for developing a light spot at the target location. The proposed WMA-GA has focused the light more tightly into a spot and converged approximately after 600 iterations with a fitness score of 1000. On the other hand, the standard GA progressed extremely slowly to a fitness score of less than 140, which was around $746\%$ less than WMA-GA. \Cref{fig:fig4} shows the initial intensity and the final focused spot images obtained with WMA-GA and standard GA in the simulation. The proposed WMA-GA has shown better background suppression and contrast advancement in a lesser number of iterations. The standard GA has performed poorly in focusing the light spot where the peak-to-background ratio did not advance even after 1000 iterations.

\subsubsection{Fitness Function and Its Impact on Structuring Light Through Scattering Media}

\begin{figure}[h]
\centering\includegraphics[width =1 \linewidth]{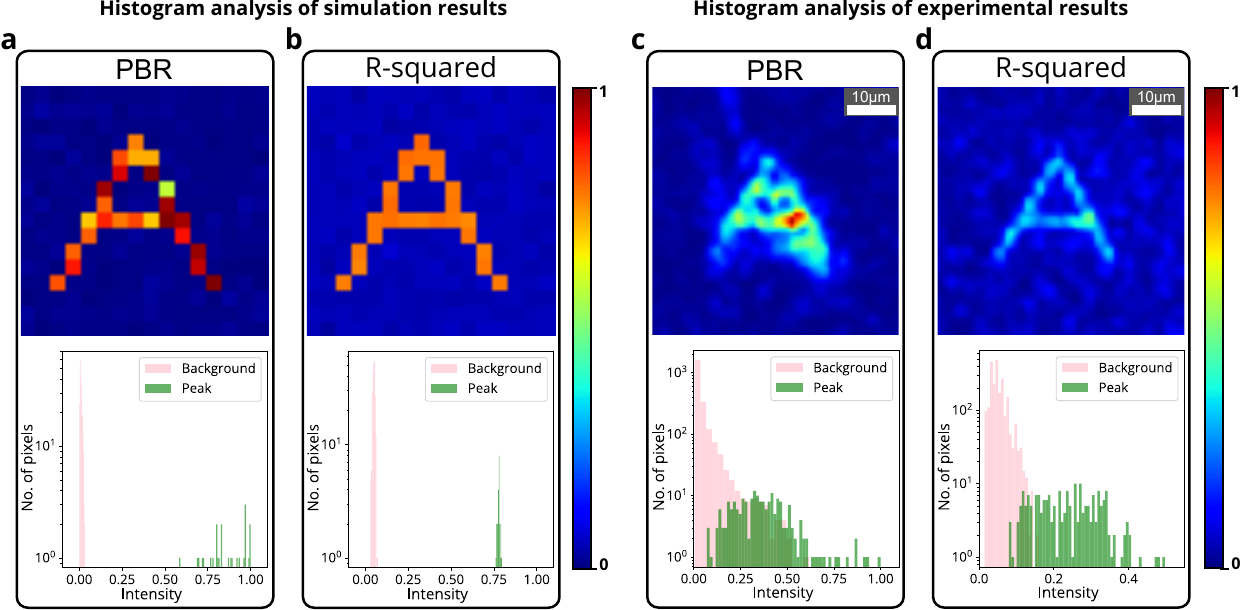}
\caption{\textbf{Comparison between R-squared metric and peak-to-background (PBR) fitness functions.} Simulation results show \textbf{a.} reconstructed structure A using PBR fitness function and its intensity histogram, \textbf{b.} reconstructed structure A using R-squared metric fitness function and its intensity histogram. Similarly, the experimental results show \textbf{c.} reconstructed structure A using PBR fitness function and its intensity histogram, \textbf{d.} reconstructed structure A using the R-squared metric fitness function and its intensity histogram.}
\label{fig:fig5}
\end{figure}

The fitness feature in iterative optimization algorithms is an important parameter that is sensitive to the desired solution. It has been observed that the proposed R-squared fitness function and the most commonly used peak-to-background ratio (PBR) fitness function perform differently based on the complexity of the structure at the region of interest (ROI). \Cref{fig:fig4} shows that a PBR fitness function has performed well while focusing a light spot at the target location. However, the PBR based fitness function has not been able to resolve and equalize the intensity for the complex structure as the target image (\Cref{fig:fig5}a--d). To solve complex structures, an even distribution of intensity across all pixels at the target location is essential. In the experiment, the PBR based fitness function completely fails to construct a complex structure like the alphabet letter \textbf{A} (see \Cref{fig:fig5}c). On the other side, R-squared metric based fitness function outperforms in constructing the structures clearly by advancing the contrast and resolution. The insets of \Cref{fig:fig5} show the histograms of the intensity distribution at the target area for both R-squared metric and PBR fitness functions. The R-squared metric is a measure of variance between two data sets \cite{steel_torrie_1960}, and it has been frequently used in machine learning and regression models \cite{Coelho2022}. The R-squared metric is calculated over M samples and measured between two sets of variables, the camera image ($I$) and the target image ($S$), as follows;

\begin{equation}
R^2 = 1 - \frac{\sum_{j=1}^M (I_j - S_j)^2} {\sum_{j=1}^M ({I_j - \overline{I}})^2}\,,\qquad \textit{where} \,\,\,\overline{I} = \frac{1}{M} \sum_{j=1}^M I_j
\label{Rsquared2}
\end{equation}

This R-squared coefficient value comes between 0 and 1. It quantifies the relationship between the movement of a dependent and an independent variable. The coefficient of 1 refers to a perfect matching among the two sets of data, and the value near 0 represents no linear relationship between the two data sets. A detailed analysis of the contrast enhancement and background noise suppression in the presence of varying noise percentages for standard GA and WMA-GA with PBR and R-squared fitness function is given in the Supporting Information (Figure S3).

\subsection{Characterization of Experimental Setup and Formation of 2D/3D Complex Structures Through Biological Tissue Media}

\begin{figure}[h!]
\centering\includegraphics[width = 0.95 \linewidth]{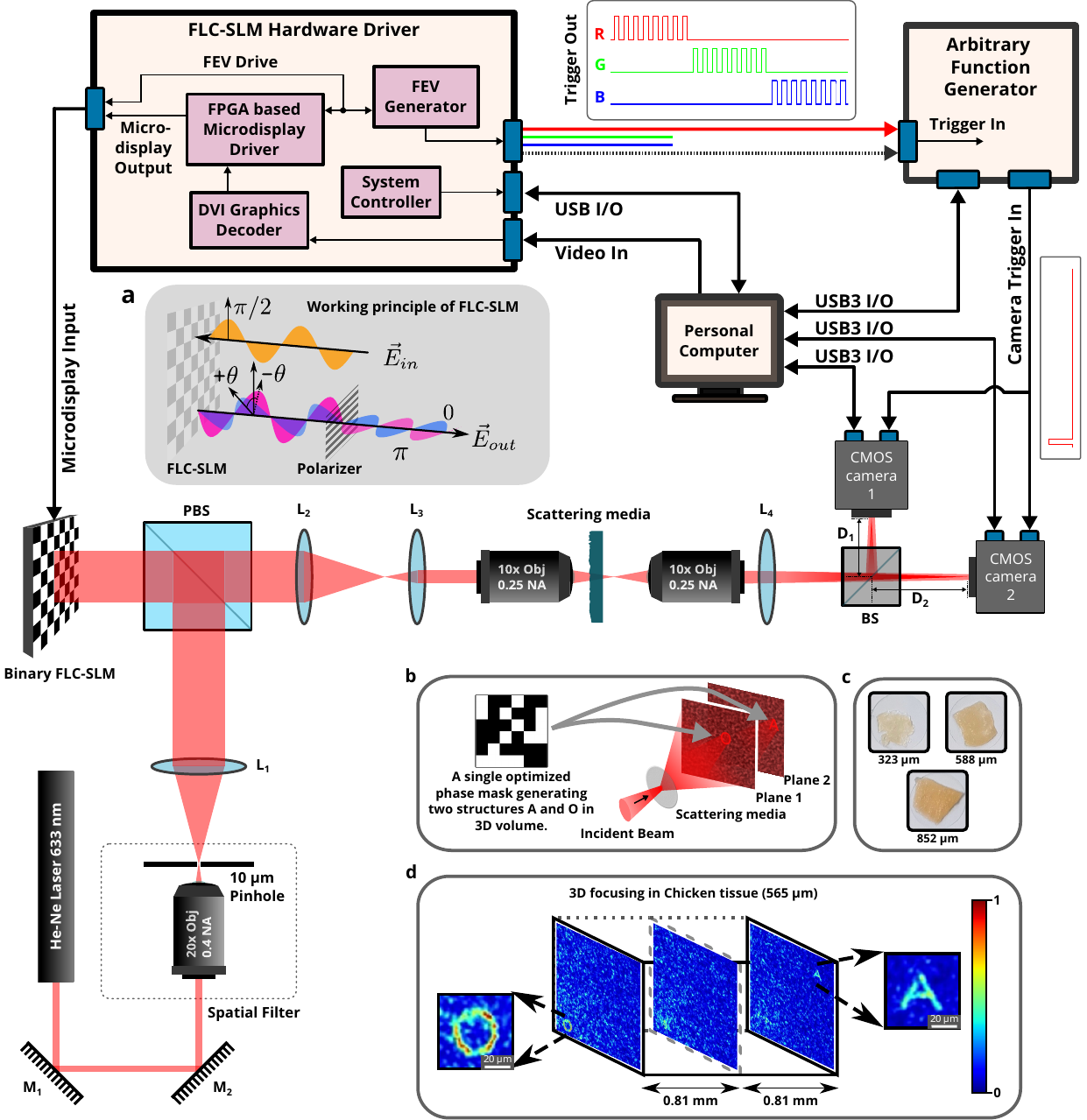}
\caption{\textbf{Schematic of the experimental setup.} Where, $\mathrm{{M_1}}$ and $\mathrm{{M_2}}$ : mirrors 1 and 2. $\mathrm{L_1}$, $\mathrm{L_2}$, $\mathrm{L_3}$ and $\mathrm{L_4}$ : lenses 1, 2, 3 and 4. PBS : polarising beamsplitter, BS : $50:50$ beamsplitter. Camera 1 and Camera 2 are placed at distances $\mathrm{D_{1}}$ and $\mathrm{D_{2}}$ from the beamsplitter respectively. Furthermore, \textbf{a.} Working principle of the FLC-SLM, \textbf{b.} Illustration for multiple complex hetero-structures formation simultaneously in 3D space using a single optimized phase mask, \textbf{c.} Chicken tissue samples of thickness 323, 588, and 852\,$\text{\textmu m}$, \textbf{d.} Experimental results for 3D complex structure formation through chicken tissue with an optimized single phase-mask using WMA-GA.}
\label{fig:fig6}
\end{figure}

The detailed schematic of the experimental system with the various hardware building blocks, experiment tissue samples, and the constructed 3D volume image is shown in \Cref{fig:fig6}. The system design consists of a master controller, i.e., the FLC-SLM hardware driver. This hardware driver is connected further with the responders, i.e., the FLC-SLM's micro-display unit and the arbitrary function generator which triggers both the cameras. The light from a He-Ne laser of wavelength 633\,nm passes through a spatial filter and falls on the SLM. Thereafter, the modulated wavefront reflected from FLC-SLM passes through a set of optical components and falls on the scattering media. To facilitate the formation of multiple complex structures \textit{simultaneously} at different depths in the 3D volume, a beam splitter is used to split the speckle field into two parts. These two parts are imaged by cameras placed at two different depths. Camera 1 is placed at distance $\mathrm{D_1}$, which has an option of position shift. Camera 2 is placed at distance $\mathrm{D_2}$ to visualize the 3D volume. Furthermore, a set of sequential hardware operation instructions are sent from the personal computer (PC) to the FLC-SLM display head and the cameras for acquiring the output speckle field generated by the tissue sample.

The simulation results have been validated with the developed experimental setup, where the proposed WMA-GA algorithm has been tested to focus complex 2D as well as 3D structures. Commercial GG diffusers with different grit sizes (120, 220, 450, and 600 grit) have been used as scattering media. Fresh chicken tissue samples of thickness 323, 588, and 852\,$\text{\textmu m}$ have been used as a biological sample for demonstration.

\begin{figure}[h!]
\centering\includegraphics[width = 0.6 \linewidth]{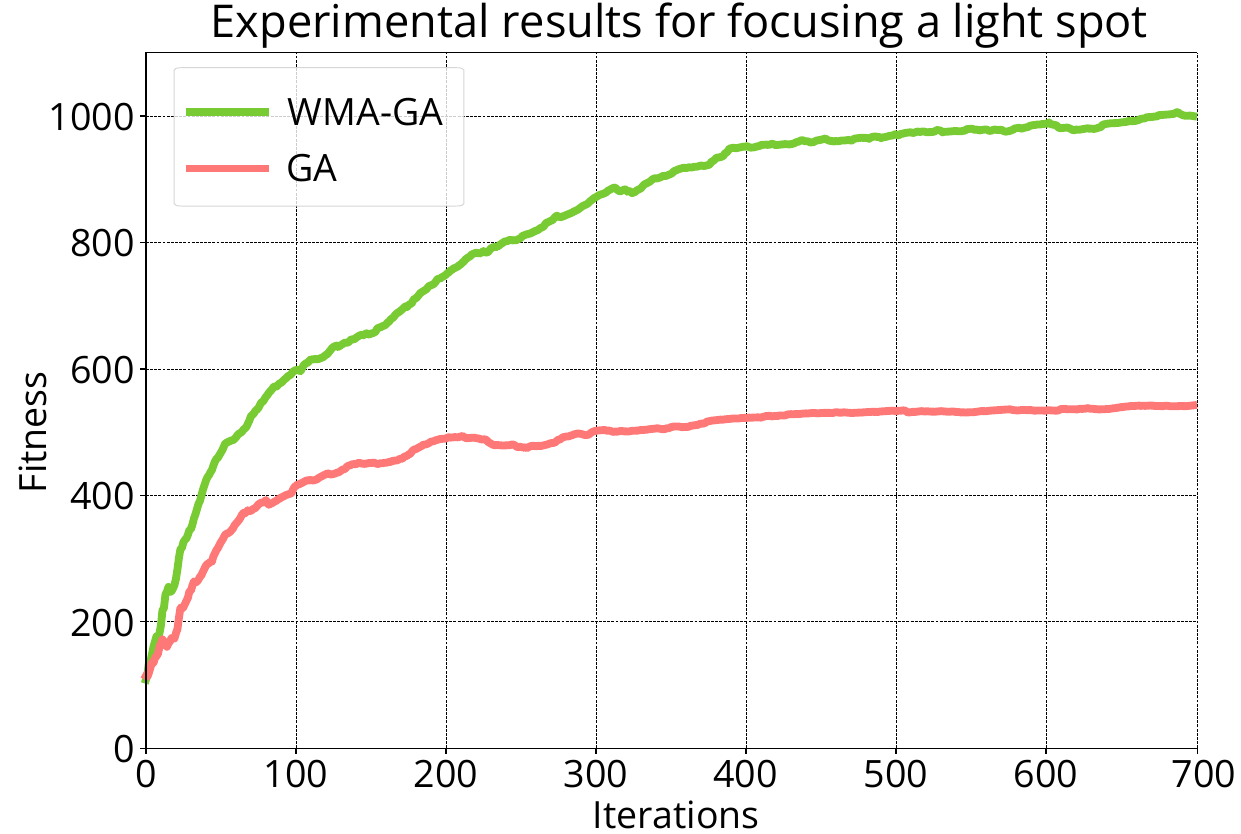}
\caption{\textbf{Experimental fitness v/s iterations plot.} The figure shows the progress of fitness score v/s number of iterations for focusing light using WMA-GA and standard GA.}
\label{fig:fig7}
\end{figure}

\begin{figure}[h!]
\centering\includegraphics[width = 0.8 \linewidth]{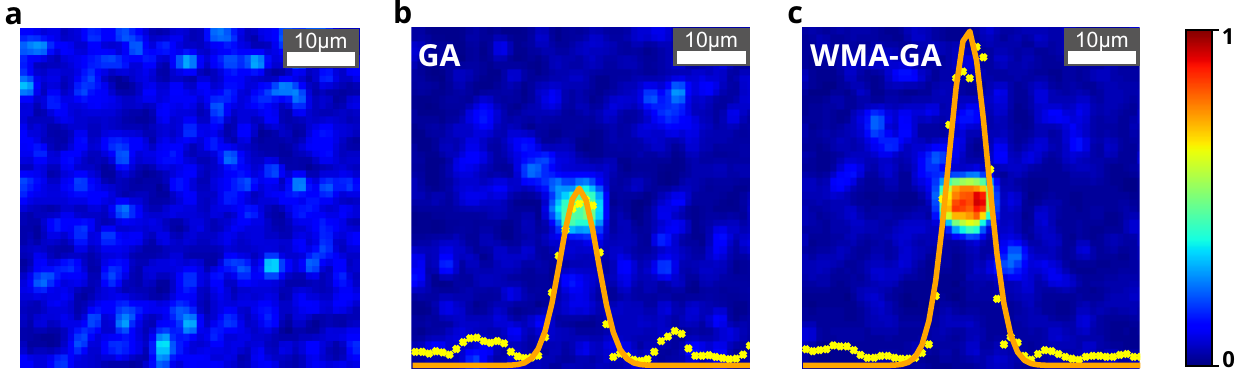}
\caption{\textbf{Experimental results for focusing a light spot.} Figure \textbf{a.} shows the initial image before focusing. Figure \textbf{b.} shows the focused spot formed by standard GA, and figure \textbf{c.} shows the focused spot formed by WMA-GA. 450 grit GG diffuser has been used as a scattering media.}
\label{fig:fig8}
\end{figure}

\begin{figure}[h!]
\centering\includegraphics[width = 1\linewidth]{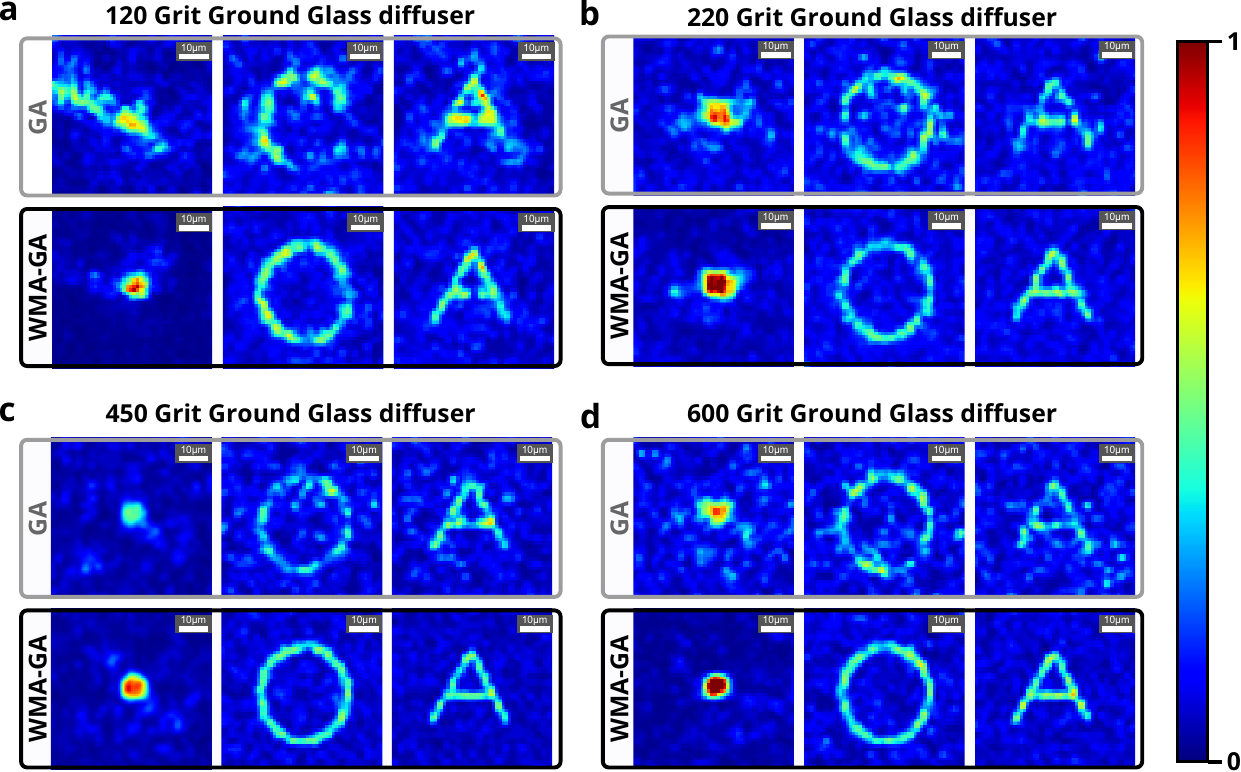}
\caption{\textbf{Experimental results for focusing light spot as well as complex structures in a 2D plane using different grit sizes of ground glass diffusers.} Comparison between WMA-GA and standard GA is shown for \textbf{a.} 120 grit, \textbf{b.} 220 grit, \textbf{c.} 450 grit and \textbf{d.} 600 grit size. All images in this figure have the same scale bar (10\,$\text{\textmu m}$).}
\label{fig:fig9}
\end{figure}

\Cref{fig:fig7} shows the progress of the fitness function for focusing a light spot after $700$ iterations using a $450$ grit GG diffuser. Here, the fitness function has been taken as PBR. The standard GA has achieved a fitness of $542$ after $700$ iterations, while WMA-GA has achieved the same in mere $75$ iterations. The maximum fitness score of $1000$ has been achieved by the WMA-GA, which is $185\%$ higher than the standard GA. \Cref{fig:fig8} shows the initial intensity and the final focused spot images along with Gaussian fit for WMA-GA and standard GA. It can be seen that WMA-GA constructed a brighter spot at the target location, which corroborates well with the simulation results. Formation of the light spot and complex patterns through GG diffusers of different grit sizes for both standard GA and WMA-GA are shown in \Cref{fig:fig9}. For a light spot, WMA-GA has outperformed in intensity enhancement as well as background suppression. It has also been observed that standard GA could not focus complex structures like the alphabet letters \textbf{A} and \textbf{O} clearly, while WMA-GA was able to form sharp \textbf{A} and \textbf{O} through all diffusers, including 120 grit which is highly scattering. \Cref{fig:fig9} equally shows that the background suppression of WMA-GA is superior to the standard GA by a significant margin.

\begin{figure}[h!]
\centering\includegraphics[width =1 \linewidth]{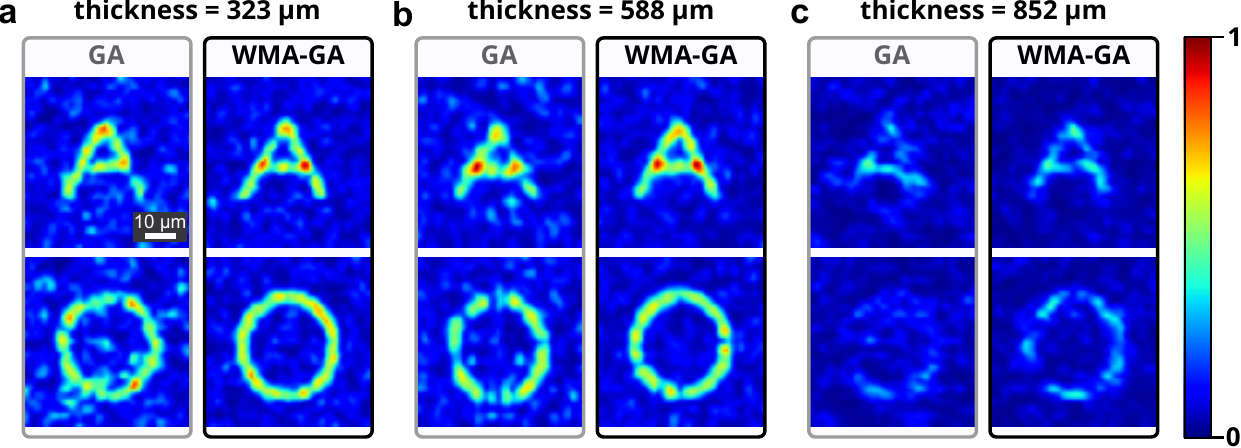}
\caption{\textbf{Experimental results for complex structure formation in the 2D space using chicken tissues of different thicknesses.} Comparison between WMA-GA and standard GA is shown for chicken tissue with thickness \textbf{a.} 323\,$\text{\textmu m}$, \textbf{b.} 588\,$\text{\textmu m}$ and \textbf{c.} 852\,$\text{\textmu m}$. All images in this figure have the same scale bar (10\,$\text{\textmu m}$).
\label{fig:fig10}}
\end{figure}

The stability of the experimental setup has been demonstrated with chicken tissue samples of thickness 323, 588, and 852\,$\text{\textmu m}$. \Cref{fig:fig10} shows the formation of complex structures through chicken tissues for the standard GA and the proposed WMA-GA. Similar to the GG diffuser results, the WMA-GA outperforms the standard GA in terms of both sharpness and background suppression for complex structure formation through the chicken tissue. As tissue sample thickness has increased to 852\,$\text{\textmu m}$, standard GA was tumbling to construct complex structures, while WMA-GA has shown better structure formation than standard GA(\Cref{fig:fig10}c). The simulation result shows that the light intensity decreases exponentially with increased tissue thickness and makes the tissue less transparent (please see Figure S18 in the Supporting Information).

\subsubsection{\textit{Simultaneous} Formation of Multiple Complex Structures in 3D Space through Tissue}

\begin{figure}[h!]
\centering\includegraphics[width =1 \linewidth]{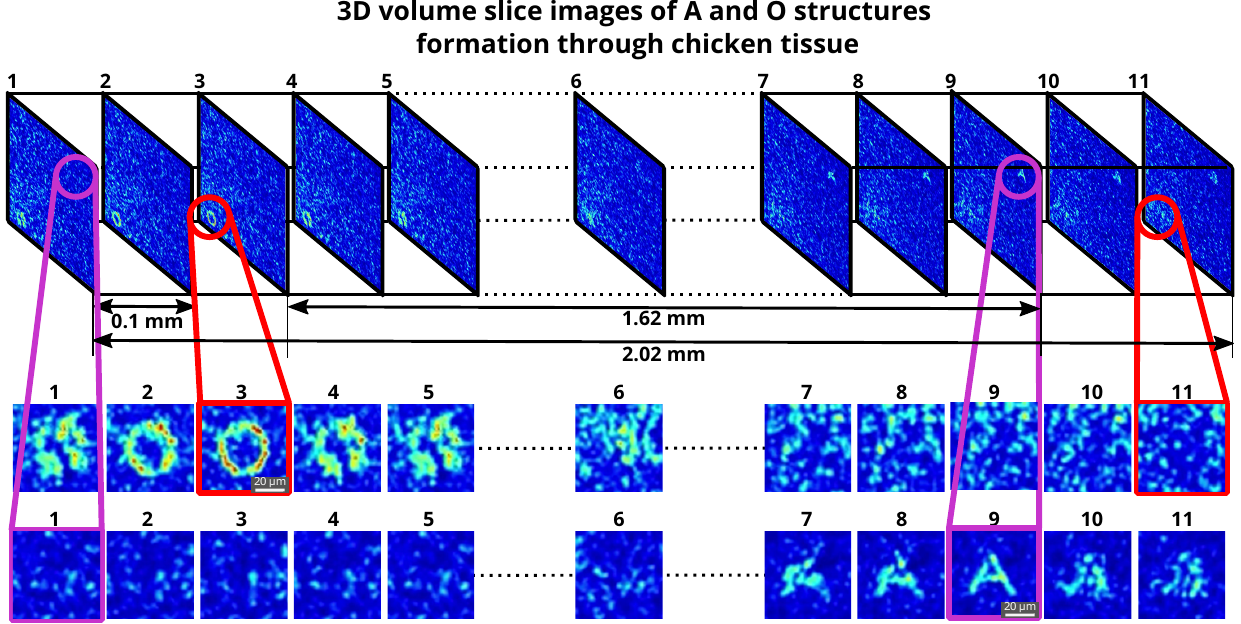}
\caption{\textbf{Experimental results for simultaneous multiple structures formation in 3D space.} 3D volume slice images of simultaneous complex structures formation through chicken tissue, where \textbf{A} and \textbf{O} structures are formed in different planes. The axial distance between highly resolved structures is 1.62\,$\text{mm}$ and lateral distance is 266\,$\text{\textmu m}$.}
\label{fig:fig11}
\end{figure}

Sequential or temporal 3D holography through scattering media using multiple phase masks has been reported previously\cite{YuNatPhot2017, Tran2019}, but \textit{simultaneous} multiple complex structure formation in 3D volume using a single phase mask has not been explored yet. An experimental system design with dual cameras have been proposed (\Cref{fig:fig6}), which simultaneously facilitates the construction of multiple complex structures in 3D space (\Cref{fig:fig11}). With the proposed method and the experimental setup, multiple complex structures have been constructed \textit{simultaneously} at different planes of 3D volume by displaying an optimized single phase-mask on the FLC-SLM. A fresh chicken tissue of thickness 565\,$\text{\textmu m}$ has been used as scattering media. An optimized phase mask has been developed using WMA-GA and R-squared cost function and displayed on the FLC-SLM for forming complex structures at multiple planes in 3D space. \Cref{fig:fig11} shows the 3D volume slice images of A and O structures formation through chicken tissue using a single optimized phase mask. The axial and lateral distances between the two complex objects have been kept 1.62\,$\text{mm}$ and 266\,$\text{\textmu m}$, respectively.

\subsection{Fluorescence Spot Formation Inside the Chicken Tissue}

\begin{figure}[h!]
\centering\includegraphics[width =0.7 \linewidth]{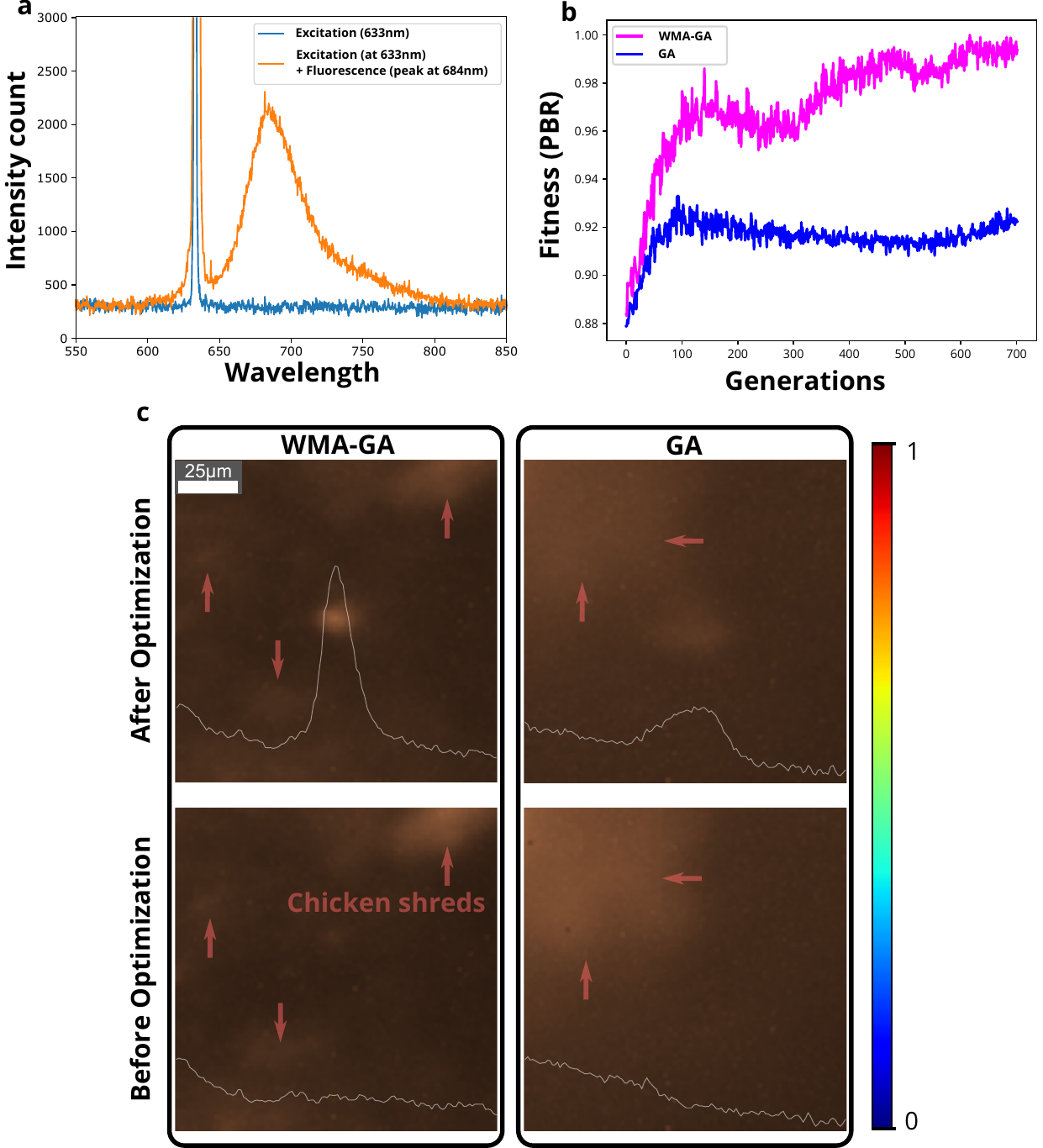}
\caption{\textbf{Fluorescence spot formation inside chicken tissue using WMA-GA in reflection mode.} Where \textbf{a.} shows the wavelength spectrum of exciting laser light and fluorescence emission photons of light together. \textbf{b.} shows the fitness score (intensity) of fluorescence emission light progress with iterations. \textbf{c.} shows fluorescence emission intensity from the targeted focus spot inside tissue for WMA-GA and standard GA after optimization and before optimization. Scale bar in the images is 25\,$\text{\textmu m}$.}
\label{fluorescence}
\end{figure}

We have demonstrated the robustness of the proposed method to control the photon-in and photon-out from the tissue media. The localized fluorescence signal extraction from the deep tissue is one of the potential applications of wavefront shaping in biomedical imaging\cite{Boniface2020, Wu2021}. The developed iterative algorithm and FLC-SLM based binary phase modulation system have been used in the reflection mode for the acquisition of fluorescence emission light in the experiment. The detailed experimental setup design for the photon-in and photon-out has been shown in the Supporting Information (Figure S17). The embedded fluorescent dye in the biological tissue has been excited by shaping the wavefront of laser light of wavelength 633\,nm. The fluorescence emission photons of light in wavelength range 680-710\,nm have been enhanced inside the tissue by the proposed WMA-GA, which works simultaneously for the photons penetrating in and out of the tissue.

An aqueous solution of Methylene blue has been used as a fluorescence material with a concentration of 40\,mg/mL. The developed system and algorithm have been demonstrated for controlling the light propagation in and out of scattering media. Results are shown in \Cref{fluorescence}. As shown in \Cref{fluorescence}c, it is visible that standard GA is not able to make tight focus spot and cannot suppress background intensity compared to WMA-GA.

\section{Conclusions and Perspectives}
Advancing contrast, increasing the efficiency of transparency for turbid media, tight focusing, fast convergence, photon transfer control, and high-resolution multiple light structures formation in 3D space have become the major demands for the wavefront shaping community to achieve. In this context, WMA-GA algorithm has been proposed to control the evolution and optimum diversity in the population, which adjusts to the problem for advancing the fitness score and convergence (\Cref{fig:fig3,fig:fig4,fig:fig5,fig:fig7,fig:fig8}). An R-squared based cost function has been introduced into the algorithm, which has outperformed in advancing the contrast and resolution (\Cref{fig:fig9,fig:fig10,fig:fig11}).

The simulation with various experimental conditions, such as different noise levels, scattering parameters, number of input modes and mutation rates, have been incorporated to mimic the experiments (please see the Supporting information Figures S2--S8 ). This analysis has demonstrated the maximum fitness score at the statistical average weighting factor W = 10.4\%, with an estimated standard deviation of 5.8. For simplicity, the weighted mutation value has been set to 10 in all simulations and experiments. Although the proposed algorithm has performed 746\% better contrast enhancement than standard GA in simulation, Whereas in the experiment, the enhancement of contrast with the proposed algorithm has been 185\% better than standard GA. Experimental noise is one of the major hindrances in feedback-based algorithms. The reduction of contrast recovery in the experiment can be attributed to various noises, such as temperature fluctuations, micron/sub-micron size dust particle movements, airflow dynamics, camera noise, and mechanical vibrations. Simulation results show that with 200\% noise, the fitness performance of WMA-GA has reduced significantly though it has shown better contrast enhancement compared to standard GA. \Cref{fig:fig7} shows that standard GA has achieved a fitness of $542$ after $700$ iterations while WMA-GA has achieved the same in mere $75$ iterations.

The developed cost-effective, calibration-free, advanced iterative wavefront shaping system and algorithm have been validated with 120, 220, 450, and 600 grit GG diffusers along with fresh \textit{ex-vivo} chicken tissue samples of thickness 323, 588, and 852\,$\text{\textmu m}$ (\Cref{fig:fig9,fig:fig10,fig:fig11}). The proposed method has demonstrated better noise tolerance and achieved a higher contrast and resolution. As an additional outcome, it has converged rapidly compared to the standard GA (\Cref{fig:fig3,fig:fig7}). As sample tissue-thickness increased to 852\,$\text{\textmu m}$, the transmission efficiency has also decreased exponentially due to the scattering and absorption of photons (supporting information Figure S18). Wavefront modulated light of wavelength 633\,nm has been focused inside the biological tissue to enhance fluorescence (680-710\,nm) emission by the proposed WMA-GA and system that works simultaneously for the photons penetrating in and out of the tissue. The designed integrated system with dual camera and R-squared cost function has constructed high-resolution multiple heterogeneous complex structures (A/O shapes) simultaneously at different depths in 3D using an optimized single phase-mask. This work has multiple potential applications, such as 3D-confocal microscopy, 3D photoacoustic microscopy and holography, photo-thermal therapy, and dosimetry. However, the efficiency of the SLM decreases as the number of multiple complex structures increases because the limited number of input modes in the optimized phase mask gets distributed to different planes.

The other advanced functionalities of the FLC-SLM, such as RGB data transfer based on three color channels and wavelength calibration-free property, always add an edge to design a new experiment. The fast pixel switching response time of ($40 \, \text{\textmu s}$) and a high refresh rate of $4.5 \, \mathrm{kHz}$ \cite{Park:20} make the FLC-SLM suitable for applications like tissue imaging, live cell imaging and photoacoustic microscopy. Despite advancements in algorithms and SLM refresh rate, the operating speed of the whole system is bottlenecked by the slow data transfer rate between the camera and the PC. However, the delay in data transfer from the camera to the PC can be reduced drastically by using a multi-channel data transfer protocol like \textit{CoaXPress}. A faster acquisition speed will further reduce the number of iterations required to reach convergence as it will nullify the noise generated due to beam shift, temperature fluctuations and the change in response of the camera sensor. The above advantages, combined with its cost-effectiveness, make the system more suitable for designing various complex wavefront shaping experiments.
\section{Methods}
\subsection{Simulation Model}
The simulation has been designed in an open-source Python 3 programming language, and NumPy has been used for processing the matrices. A matrix of dimensions $200\times200$ has been taken as the input mode from the FLC-SLM, which corresponds to $N = 40,000$ input modes ($\vec{E}_{in}$). Another matrix of dimensions $16\times16$ has been considered as the output wavefront, corresponding to $M=256$ output modes ($\vec{E}_{out}$). The transmission matrix $T$ of dimensions $M\times N$ has been generated using a complex Gaussian random distribution ($\mu_T = 0$ and $\sigma_T = 0.1$) to mimic the scattering of light. Further, a 30 \% noise $\delta$ has been added to the output mode intensity to mimic the experimental conditions (\Cref{eq3}).

In the beginning of the algorithm, a population ($P$) of random binary phase masks has been generated using a discrete uniform distribution of values 0 and 255, which correspond to 0 and $\pi$ phase, respectively. Thereafter, WMA-GA has evolved the population to produce the desired result at the output mode (\Cref{fig:flowchart_fig}). A population size of 200 has been chosen as it provides a good trade-off between speed and enhancement. Two parents $\vec{P}_i$ and $\vec{P}_j$ have been selected with a biased probability towards a higher fitness score. The descending order of the phase masks has been ranked according to their fitness score, which has used later for parents selection. The crossover rate ($r_c$) has been kept at the standard value of 50\%. The initial mutation rate has been fixed at 1\%, which exponentially decays to 0.5\% with a constant decay rate of $\lambda = 500$. A typical choice of mutation rate in genetic algorithms is below 5\% \cite{doi:10.1137/S009753979732565X}. Unlike in the case of NLC-SLM, binary FLC-SLM opts for only two phase values, 0 and $\pi$, which implies that a mutation rate ($r_m$) of 99\% is the same as 1\%, because 0 and $\pi$ can be redefined by changing the reference. This mathematical calculation has been provided in the Supporting Information (eqs S2--S6).

\subsection{Experimental System Design with FLC-SLM for Proposed WMA-GA}

\begin{figure}[h!]
\centering\includegraphics[width = 0.95 \linewidth]{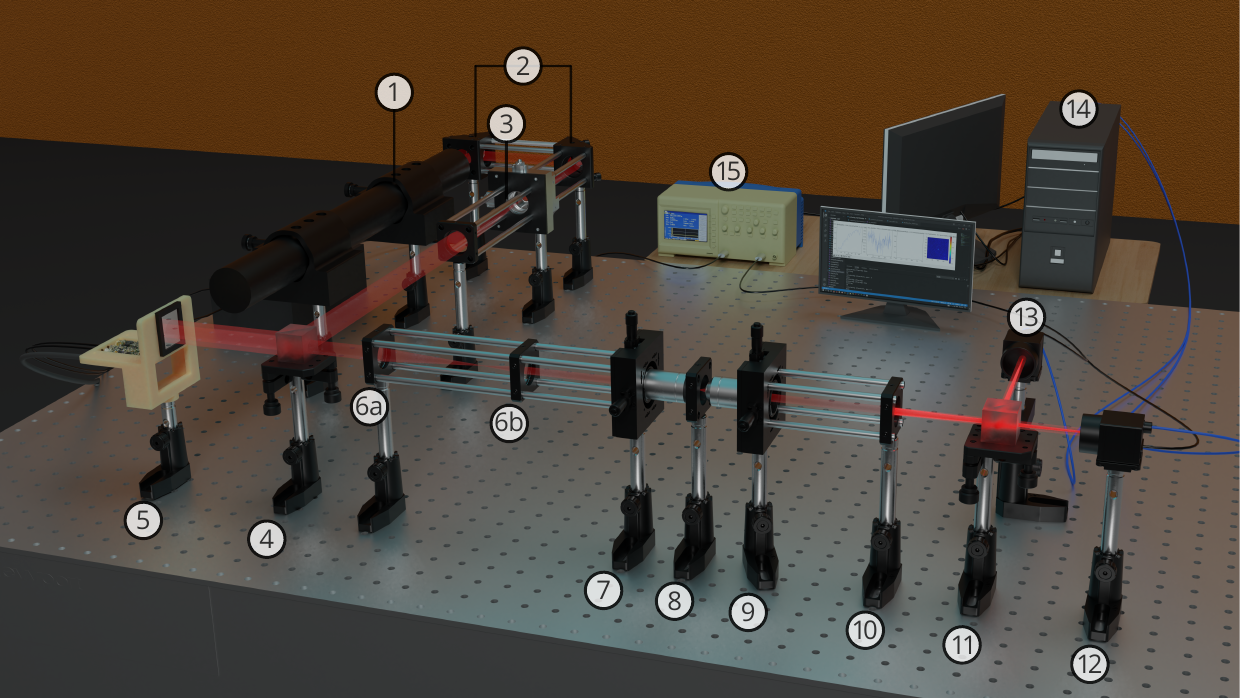}
\caption{\textbf{3D schematic of experimental setup.} Components: 1. He-Ne laser, 2. Mirrors $\mathrm{M_1}$ \& $\mathrm{M_2}$, 3. Spatial filter, 4. Polarizing beam splitter (PBS), 5. FLC-SLM, 6a-6b. 4F setup, 7. $1^{st}$ objective, 8. Scattering media, 9. $2^{nd}$ objective, 10. Lens ($f = 50\, \text{mm}$), 11. $50:50$ Beam splitter, 12. CMOS camera-1, 13. CMOS camera-2, 14. PC, and 15. Arbitrary function generator.}
\label{fig:fig13}
\end{figure}
A 12\,mW He-Ne Laser (633\,nm, Newport) consisting of vertical linearly polarized light with a polarization ratio of $500:1$ is used in the built system. Two flat mirrors, $\mathrm{M_1}$ and $\mathrm{M_2}$ are used for laser beam alignment. Along the path, a spatial filter system (Thorlabs, KT311/M) is placed consisting of a pinhole ($\phi=10\, \text{\textmu m})$ and objective (20X, Numerical Aperture (NA) = 0.40) for eliminating the higher-order noise from the beam. Thereafter, the spatially filtered diverging beam is collimated by a lens $\mathrm{L_1}$ $(f = 250 \, \text{mm})$ to get a pure flat beam profile on the surface of FLC-SLM. A polarising beam splitter (PBS) and FLC-SLM (ForthDD, SXGA-R5) are used for wavefront modulation. The modulated wavefront is passed through a $4F$ setup and entered into an objective(10X, $\text{NA} = 0.25$) which transmits the wavefront through the scattering media. The power of the incident beam just before the tissue sample has been measured and found to be 0.74\,mW. A second objective (10X, $\text{NA} = 0.25$) is placed behind the scattering media. For the simultaneous construction of multiple complex structures in 3D volume, the camera $\mathrm{D_1}$ (Thorlabs, DCC3260C) is placed to acquire image at different depths in 3D space. The camera $\mathrm{D_2}$ (Basler acA800-510uc) is used to capture the image and later make the feedback signal for algorithms. The camera $\mathrm{D_1}$ has the option to move back and forth to construct multiple complex structures at different depths.

The signal from the PC to the SLM driver module is sent via a video card. Each image is a combination of 24 bit-planes, i.e., 24 bit information per pixel and 8 bit per channel (RGB). The hardware module of the SLM splits the RGB signal into 24 single-bit black and white images. These 24 single-bit images are sent and displayed on the SLM screen sequentially. So conclusively, a total of $24\times 60 = 1440$ binary images are displayed on the SLM screen in 1 second. Each bit plane is displayed on the SLM screen for a duration of 219.02\,$\text{\textmu s}$. The hardware driver module of the SLM is programmed to generate an output electrical signal which becomes high or low in synchronization with the display of each bit plane. This signal is passed on to the function generator to generate a new signal with $+3\,\mathrm{V}$ to trigger the two CMOS cameras. The other advanced features of the FLC-SLM, such as three color channels, are made available to use in combined form or independently. Multiple camera interfacing has been carried out using the additional controller-responder signal generator for constructing simultaneous multiple heterogeneous complex holography in 3D space.\\

\subsection{Preparation of Chicken Tissue Samples for Experiment}
A fresh chicken (weight = 2.62\,$\text{kg}$, age = 10 weeks, measured density = 0.92\,$\mathrm{g/cm^3}$) has been procured from the local market. The whole thigh of the chicken has been kept in the fridge for 4 hours at a constant temperature of $-14 \, \mathrm{^{\circ}C}$ to facilitate the slicing. A sterilized surgical knife has been used to cut the chicken muscles into several slices. The measured thickness of the sliced chicken tissues came out to be 323, 588, and 852\,$\text{\textmu m}$. The sliced chicken muscle has been sandwiched between two microscope glass cover slips. A drop of Glycerin has been used to preserve the sample and prevent it from drying.


\section*{Associated Content}
\begin{suppinfo}
The Supporting Information is available free of charge at DOI: 

The document contains details about the following: The data of the extensive study on the impact of various experimental conditions; a detailed analysis of the contrast enhancement and background noise suppression in the presence of varying noise percentages for standard GA and WMA-GA with PBR and R-squared fitness function; mathematical calculation of output modes in binary phase FLC-SLM and derivation for mutation rate in the case of binary masks; construction of word IISER in the experiment; the process of synchronizing the cameras with the FLC-SLM; microscopic images of GG diffusers;
measurement of camera zoom factor; chicken tissue samples used in the experiment; intensity line-plot and histogram analysis for focused spot; schematic setup for fluorescence spot formation; intensity drop v/s distance analysis inside scattering media; and experimental noise analysis (PDF).
\end{suppinfo}

\section*{Author Information}
\textbf{Corresponding Author}\\
*E-mail: skbiswas@iisermohali.ac.in\\
\textbf{ORCID}\\
Amit Kumar: orcid.org/0000-0002-7413-850X\\
Sarvesh Thakur: orcid.org/0000-0003-0601-6583\\
S.K. Biswas: orcid.org/0000-0003-2087-8112\\

\section*{Competing interests}
The authors declare no competing interest.

\begin{acknowledgement}
 The authors are thankful to IISER Mohali startup fund and IMPRINT funding agency for their support. The authors are thankful to the maintainers of the website wavefrontshaping.net for providing a comprehensive guide to the field of wavefront shaping.
\end{acknowledgement}

\bibliography{achemso-demo}

\end{document}